\documentclass{aa}  

\usepackage{graphicx}
\usepackage{txfonts}
\usepackage{lipsum}
\usepackage{subcaption}         
\usepackage{lscape}             
\usepackage{placeins}           

\usepackage[version=4]{mhchem}
\usepackage{xcolor}

\usepackage{natbib}
\bibpunct{(}{)}{;}{a}{}{,} 

\usepackage{txfonts}
\usepackage[pdftex,plainpages=false,
    pdfpagelabels,
    pdfusetitle,  
    bookmarksnumbered,
    linkbordercolor={1 1 1},
    hidelinks=true,
    colorlinks={ true },
]{hyperref}

\newcommand{\rateU}{\ensuremath{\text{cm}^{3}\,\text{s}^{-1}}}
\newcommand{\kcalmol}{\ensuremath{\text{kcal}\,\text{mol}^{-1}}}

\newcommand{\massU}{\ensuremath{m/z}}

\newcommand{\HCOp}{\ce{HCO+}}
\newcommand{\HOCp}{\ce{HOC+}}

\newcommand{\iNN}{\ce{^{15}N2}}

\newcommand{\iNNHp}{\ce{^{15}N2H+}}
\newcommand{\HHHp}{\ce{H3+}}

\newcommand{\COp}{\ce{CO+}}

\newcommand{\HH}{\ce{H2}}

\begin{document}

   \title{Isomer-resolved ion-molecule reaction kinetics of \ce{HCO+} and \ce{HOC+}}

%
%
%

   \author{M. Jim\'{e}nez-Redondo\inst{1}\email{mjimenez@mpe.mpg.de}        
        \and O. Sipil\"{a}\inst{1}\email{osipila@mpe.mpg.de}
        \and P. Dohnal\inst{2}\email{petr.dohnal@mff.cuni.cz}
        \and R. Dahl\inst{1,3}\email{dahl@uni-bonn.de}
        \and P. Caselli\inst{1}\email{caselli@mpe.mpg.de}
        \and P. Jusko\inst{1}\corrauth{pjusko@mpe.mpg.de}
        }

   \institute{Max Planck Institute for Extraterrestrial Physics, 
Gießenbachstraße 1,
85748 Garching, Germany
   \and Department of Surface and Plasma Science, Faculty of Mathematics and Physics, 
Charles University, V Hole{\v s}ovi{\v c}k{\' a}ch 2, 18000 Prague, Czech Republic
   \and Mulliken Center for Theoretical Chemistry, University of Bonn,
Beringstraße 4,
53115 Bonn, Germany}

   \date{Received September 30, 20XX}

  \abstract
{The \HCOp/\HOCp\ abundance ratio has been observed to vary significantly in different astrophysical environments,
making it an important diagnostic tool for the interstellar medium.
However, uncertainties in the reaction rates and branching ratios of key chemical reactions involving these ions
affect the conclusions drawn from observations of either these species or other species that derive from them.
Here, we present isomer-resolved ion-molecule reaction rate coefficients of reactions involving \HCOp\ and \HOCp\ using a cryogenic ion trap experiment.
We measure the rate coefficients of the isomerization reaction of \HOCp\ to \HCOp\ with \ce{H2} and \ce{CO},
as well as three different reactions producing both isomers, namely \ce{CO+} + \ce{H2}, \ce{H3+} + \ce{CO}, and \ce{ArH+} + \ce{CO}.
Our results show that the reaction of \COp\ with \ce{H2} predominantly produces \HCOp\ with a negligible yield of \HOCp,
in agreement with recent theoretical predictions but in major contrast to the generally accepted branching ratio based on previous experimental results.
We also find that the isomerization reaction of \HOCp\ to \HCOp\ is efficient with both \ce{H2} and \ce{CO}.
We assess the impact of our new measured reaction rate coefficients on the predicted \HCOp/\HOCp\ abundance ratio using a state-of-the-art astrochemical model.}


   \maketitle

\nolinenumbers

\section{Introduction}

The formyl cation, \HCOp, and its higher-energy constitutional isomer, the isoformyl cation \HOCp,  
are among the simplest isomeric systems and, as such, play an important role across many scientific fields 
including astronomy. 
Indeed, \HCOp\ was detected more than 50 years 
ago in several interstellar sources as one of the first molecular ions, initially under the name ``X-ogen'' \citep{Buhl1970}, 
prior to its laboratory identification by \citet{Woods1975}.
About ten years later, \HOCp\ was first detected in the laboratory by microwave spectroscopy \citep{Gudeman1982}, 
followed shortly afterward by its detection in the interstellar medium \citep{Woods1983}.

Measurements of the [\HCOp]/[\HOCp] abundance ratio in astrophysical sources reveal large variations
driven by different environmental conditions. For example, in dense molecular clouds typical [\HCOp]/[\HOCp] values 
are found between about 1000 and 9000 \citep[e.g.,][]{Ziurys1995,Apponi1997}, while in photodissociation 
regions (PDRs) and diffuse clouds the ratio drops to $\sim 50-120$ \citep[e.g.,][]{Fuente2003,Fuente2008,Liszt2004}. 
Similar low values have also been measured toward the starburst galaxy NGC253, where enhanced radiation 
fields and/or cosmic-ray ionization rates have been proposed \citep{Harada2021}. 
The [\HCOp]/[\HOCp] ratio is thus an important diagnostic tool for several phases 
of the interstellar medium, from diffuse clouds, to molecular clouds, to star-forming regions, to 
planet-forming disks and finally to galaxies. For this reason, several theoretical and laboratory studies have been carried 
out to understand the observed behavior. Ab initio calculations of \cite{Herbst1996} revealed that 
the isomerization reaction
\begin{equation}
    \HOCp + \HH \to \HCOp + \HH
\end{equation}
involves a barrier, causing the low-temperature rate coefficient to decrease substantially below about 50 K, thus improving agreement with observations of 
\HOCp\ in dark clouds. The larger [\HCOp]/[\HOCp] ratios in dense clouds compared to diffuse clouds could 
be attributed to the fact that isomerization of \HOCp\ is much faster than dissociative recombination with electrons \citep{Kassayova2025} when 
the electron fraction is low ($x(e)$ $\simeq$ 10$^{-7}$ vs. 10$^{-4}$), as discussed in \cite{Liszt2004}. 

Unfortunately, there are still uncertainties in the rate coefficients and branching ratios of important 
chemical reactions involving \HCOp\ and \HOCp, thus precluding a quantitative use of the [\HCOp]/[\HOCp] 
ratio as an astrophysical tool. In this work, we present ion trap measurements of the rate coefficients 
for isomerization reactions of \HOCp\ to \HCOp\ and for the \ce{CO+ + H2} reaction, long believed to 
produce \HCOp\ and \HOCp\ with a branching ratio of 1:1 \citep{Freeman1987b, Smith2002}. Our results show that this reaction does not 
produce significant amounts of \HOCp, in agreement with recent theoretical calculations \citep{Ramazani2009,Xiang2022}. 
We also measure other associated reaction rate coefficients and include the new values in our state-of-the-art astrochemical 
model to assess their effects on the predicted [\HCOp]/[\HOCp] ratio.

\section{Results}

\begin{figure}[ht!]
\includegraphics[width=.98\linewidth]{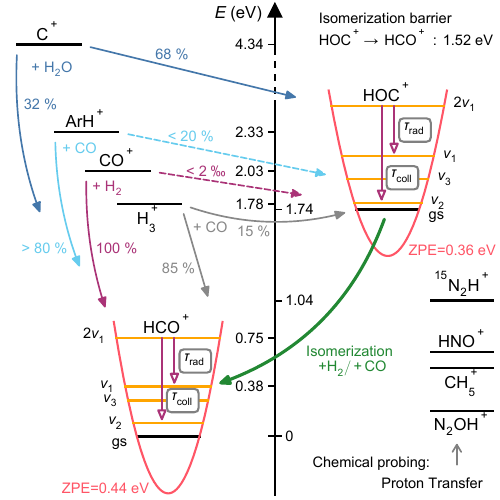}
\caption{Energy levels and possible reaction pathways leading to \ce{[H,C,O]+}. The different reactants used
for the production of both isomers are shown on the left. Branching fractions are indicated next to the arrows, with values from \citet{Yang2021}, in the case of \ce{C+ + H2O}, and this work. Viable chemical probes are shown on the right. Note that only \iNN\ was used as a chemical probe in this work. Zero-point energies \citep[ZPE, from][]{Koput2019} and selected vibrational levels ($\nu_i$) are plotted for \HCOp\ and \HOCp. $\tau_\text{rad}$ and $\tau_\text{coll}$ represent the radiative and collisional lifetimes, respectively, of the vibrationally excited \HOCp\ states.
The energy difference between the isomers is 1.74 eV \citep[from proton affinities at O and C, as reported by][]{Freeman1987b}. The
isomerization barrier from $\HOCp \to \HCOp$ of 1.52 eV is taken from \citet{Chalk1997}. Proton affinities and ionization potentials are taken from NIST \citep{Lindstrom2023}.}
\label{fig_levels}
\end{figure}

We have studied a variety of reactions involving \HCOp\ and \HOCp, including isomerization reactions and reactions producing both isomers.
Reactions of \ce{CO+ + H2}, \ce{H3+ + CO}, and \ce{ArH+ + CO}, all leading to \ce{[H,C,O]+}, are discussed in
sections~\ref{sec:cop}, \ref{sec:hhhp}, and \ref{sec:arhp}, respectively. 
We then present the results of our investigation of the isomerization reaction rate coefficient of 
$\HOCp \to \HCOp$ with \ce{H2} and \ce{CO} in section~\ref{sec:reiso}. 
A schematic view of the processes studied, along with the relevant energy levels,
is depicted in Fig.~\ref{fig_levels}. 
Branching ratios for the reactions of interest are shown in the figure and are also collected 
along with the derived rate coefficients in Tab.~\ref{t_results} at the end of this section.

\subsection{\ce{CO+} + \ce{H2}}\label{sec:cop}

Historically, the reaction $\ce{CO+} + \ce{H2}$ was often assumed to produce \HCOp\ and \HOCp\ in comparable amounts. 
This view is mainly based on the 1:1 branching ratio reported by \citet{Freeman1987b} and later supported by \citet{Smith2002}, 
and it is still reflected in current astrochemical networks \citep[e.g.,][]{Wakelam2024}.

However, more recent work points to a very different picture. 
Trajectory calculations and higher-level theory predict that the \HOCp\ channel is strongly suppressed, especially at low collision energies. \citet{Ramazani2009} claim that
\emph{``not a single trajectory was observed to enter the \ce{HOC+} + \ce{H} product valley''},
indicating negligible \ce{HOC+} yield. More recent calculations by \citet{Xiang2022} reach the same conclusion, stating that
\emph{``The product \HOCp\ can be formed by three reaction paths:
two of them have a barrier of 7.16 \kcalmol\ (TS5) and 15.56 \kcalmol\ (TS6), respectively, and the other has no
barrier by going through trans-\ce{HCOH+} but requires a complicated configuration change''} and continues:
\emph{``According to our QCT calculations, the reaction \ce{H2} + \ce{CO+} $\to$ \ce{H} + \HOCp\
virtually cannot take place at low collision energies below $5\;\kcalmol$ by going through the barrierless reaction path.''} Other experimental and combined experimental-theoretical studies either support \citep{Dietl2013} or their results are not incompatible \citep{Rosati2007} with dominant \HCOp\ formation.
Moreover, no \HOCp\ transitions could be observed
in our previous overtone spectroscopic study \citep{Jimenez2024},
when we attempted to use the \COp\ + \HH\ reaction as a source of \HOCp.
This disagreement in the literature motivates a direct, isomer-resolved measurement of the product branching ratio.


The reaction of \ce{CO+} with \ce{H2} is highly exothermic \citep{Ramazani2009}:
\begin{align}
\ce{CO+} + \ce{H2}   &\to \ce{HCO+} + \ce{H}, ~~~~~\Delta E = -2.01~\mathrm{eV} \label{r_COplusH2} \\
                     &\to \ce{HOC+} + \ce{H}, ~~~~~\Delta E = -0.32~\mathrm{eV} \nonumber
\end{align}

The resulting products can therefore carry a large amount of internal energy, especially \HCOp. Both \HCOp\ (if sufficiently excited) and \HOCp\ can react with \iNN\ to form \iNNHp.

\begin{figure}[ht!]
\includegraphics[width=.98\linewidth]{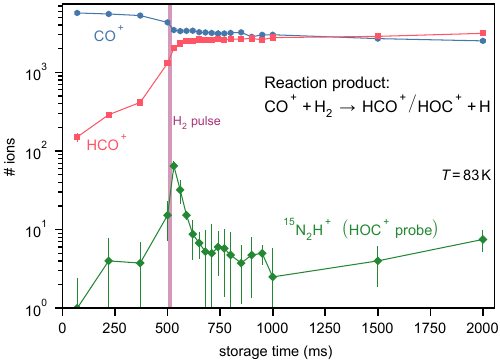}
\caption{Time evolution of the trapped ion signal for the study of the \ce{CO+} + \ce{H2} reaction in the pulsed \ce{H2} configuration (\iNN\ pulsed 50~ms before ion extraction).
    \iNNHp\ signal remains low, indicating that the \HOCp\ channel is negligible. The observed peak around the \ce{H2} injection time can be attributed to vibrationally excited \HCOp.}
\label{fig_cop}
\end{figure}

To obtain a reliable measurement of the branching ratio, the products of reaction (\ref{r_COplusH2}) must be allowed sufficient time to radiatively de-excite before being probed by \iNN.
This was achieved in two different ways (see Sec. S2.1 in the SI): 1) by using a constant (very low) \HH\ number density and pulsed probing gas; and 2) by pulsing both \HH\ and the probing gas with a variable delay in between. The data obtained with the latter method are shown in Fig.~\ref{fig_cop}.
All the methods point to a consistent result: the branching fraction for the \HOCp\ channel is negligible (below 0.002), in contrast to the 1:1 ratio assumed in current astrochemical networks.

\subsection{\ce{H3+} + \ce{CO}}\label{sec:hhhp}

 This reaction is considered to be one of the most important drivers of \ce{HCO+} content in molecular clouds \citep{Panessa2023}. It is barely exothermic for the \HOCp\ channel, and several theoretical studies predict a product ratio dominated by \HCOp.
The most recent work by \citet{Saito2021} predicts a \HOCp\ yield of approximately 17\% at 50~K and 33\% at 200~K, with similar results from earlier works \citep{Klippenstein2010, Yu2009, Le2010}.
Early experimental work by \citet{Illies1983} reported a 6\% \HOCp\ yield for the reaction, later characterized as a lower limit by \citet{Wagner-Redeker1985} after accounting for \HH-catalyzed isomerization. More recently, \citet{Carrascosa2016} reported an upper limit of 24\% for the \HOCp\ yield at 1.8~eV in a crossed beam experiment, declining to below 10\% at lower collision energies. Later theoretical works inspired by the latter study \citep{Naz2018, Zhu2019, Zhu2020} suggest a somewhat higher \HOCp\ yield (20--35\%) at lower energies.

We measured the branching ratio by pulsing \ce{CO} into the trap containing stored \ce{H3+}, monitoring mass-29$\;\massU$
product ions after \iNN\ chemical probing to identify \HCOp, since \ce{H3+} also reacts with \iNN\ to form \iNNHp\ (see Sec.~2.2 in the SI).
We obtain a branching fraction of 0.15 $\pm$ 0.04 for the \HOCp\ channel at 125~K, consistent with the lower-energy theoretical predictions and close to the values reported by \citet{Carrascosa2016}.

\subsection{\ce{ArH+} + \ce{CO}}\label{sec:arhp}

The reaction of \ce{ArH+} with \ce{CO} has been studied by \citet{Bastian2019} in a crossed beam experiment complemented by theoretical calculations. They report a branching ratio of 30\% for the \HOCp\ channel, but this value is based solely on theoretical predictions, as the experiment could not distinguish between the two isomers. Our measurements did not detect significant \HOCp\ formation; however, an accurate determination of the branching fraction was not possible because the probing gas \iNN\ also reacts with \ce{ArH+}, and the analysis is further complicated by the concurrent isomerization reaction of \HOCp\ with \ce{CO}. Given these constraints, we conservatively estimate an upper limit for the \HOCp\ branching fraction of 20\,\%. Further details can be found in Sec.~2.3 of the SI.

\subsection{Isomerization}\label{sec:reiso}

The isomerization of \HOCp\ to \HCOp\ has been studied extensively because it directly controls the abundance of the metastable isomer in cold interstellar environments. Early ion-cyclotron resonance experiments by \citet{Wagner-Redeker1985} showed that \HOCp\ reacts efficiently with \ce{D2} through a catalytic proton-transfer pathway. Subsequent selected ion flow tube (SIFT) measurements by \citet{Freeman1987b} and variable-temperature ion-trap experiments by \citet{Smith2002} established that \HOCp\ is converted to \HCOp\ rapidly by \HH\ and \ce{CO} at low temperatures, with \citet{Smith2002} reporting a rate coefficient of about $3.8 \times 10^{-10}\;\rateU$ at 25 K. 
On the theory side, the early work by \citet{Herbst1996} found that the rate coefficient may be significantly lower below about 100 K once tunneling and the detailed topology of the potential energy surface are included. More recent theoretical analyses by \citet{Vogt2009} and guided-ion-beam experiments by \citet{Solem2024} further support a hydrogen-assisted proton-transfer mechanism in which neutral collision partners promote the structural rearrangement. 

There are, however, some issues with the previous experimental studies. Both the SIFT work of \citet{Freeman1987b} and the ion trap study of \citet{Smith2002} used the reaction of \COp\ with \ce{H2} to produce the \HOCp\ used in their isomerization experiments.
They observed this reaction to yield a 1:1 mixture of \HCOp\ and \HOCp\ based on chemical probing. However, as discussed in Sec.~\ref{sec:cop}, our new measurements show that this reaction produces almost exclusively \HCOp, with negligible \HOCp\ formation at low collisional energies, an issue that would be particularly relevant for the ion trap experiment. Perhaps related to this, the chemical probing gases used in these studies, \ce{N2O} \citep{Freeman1987b}, and \ce{CH4} and \ce{NO} \citep{Smith2002}, have higher proton affinities than \iNN, meaning they are more sensitive to vibrationally excited \HCOp\ than our current study (see Sec.~1.4 in the SI). For these reasons, we have re-measured the isomerization reaction rate coefficient of \HOCp\ to \HCOp\ using \iNN\ as a chemical probe and \HOCp\ produced directly from \ce{C+ + H2O}. 

\begin{figure}[ht]
\includegraphics[width=.98\linewidth]{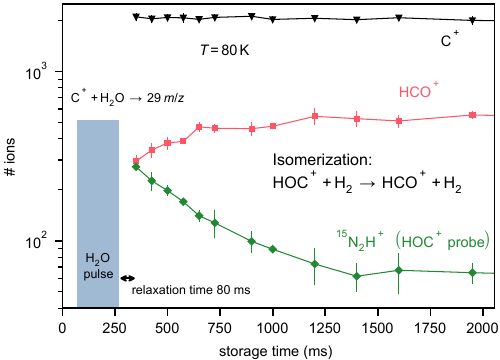}
\caption{Number of ions as a function of trapping time for the experimental determination of the isomerization rate coefficient of \HOCp\ to \HCOp\ by \HH. \HOCp\ is produced from \ce{C+} and subsequently converted to \HCOp\ in collisions with \HH. Pulsed injection of \iNN\ helps discriminate between the two isomers.}
\label{fig_iso}
\end{figure}

The measurements were performed with a constant number density of the reactant (\HH\ or \ce{CO}) and a pulsed injection of the chemical probe \iNN.
\HOCp\ was produced in situ from \ce{C+} ions using a short pulse of \ce{H2O} at the beginning of each trapping cycle. An example of the time evolution of the trapped ion signal for the isomerization reaction with \HH\ is shown in Fig.~\ref{fig_iso}. 

The isomerization rate coefficients measured in this work are $(4 \pm 1) \times 10^{-10}\;\rateU$ for the reaction with \HH\ at 80~K and $(9 \pm 2) \times 10^{-10}\;\rateU$ for the reaction with \ce{CO} at 125~K. The value for \HH\ is actually in very good agreement with the previous measurements by \citet{Freeman1987b} and \citet{Smith2002} despite the issues discussed above. The reason for this agreement may be fortuitous due to the quenching rate coefficient for vibrationally excited \HCOp\ by \HH\ also having a similar value \citep{Jimenez2024}. The rate coefficient obtained for the isomerization reaction with \ce{CO} is somewhat higher than the value reported by \citet{Freeman1987b}, but still within the error bars.

\begin{table}[ht]
  \caption{Reaction rate coefficients $k$ and branching fractions for reactions producing and destroying \HCOp\ and \HOCp. 
  }\label{t_results}
  \begin{center}
  \begin{tabular}{llccc}
    \multicolumn{2}{l}{Reaction}       & $k$     & Fraction        \\
    \hline
    $\HOCp + \HH$      & $\to \HCOp + \HH$     & $4(1)$   &  $-$           \\
    $\HOCp + \ce{CO}$  & $\to \HCOp + \ce{CO}$ & $9(2)$   &  $-$               \\
    \hline\hline
    $\COp + \HH$       & $\to \HCOp + \ce{H}$  & $15(2)^a$   & $\approx 1$           \\
                       & $\to \HOCp + \ce{H}$  &          & $<0.002$             \\
    \hline
    $\HHHp + \ce{CO}$  & $\to \HCOp + \HH$     & 20(2)$^{b}$       & 0.85(4)         \\
                       & $\to \HOCp + \HH$     &          & 0.15(4)            \\
    \hline
    $\ce{ArH+}+\ce{CO}$& $\to \HCOp + \ce{Ar}$ & $8.2^{\text{Lang.}}$    & $>0.8$         \\
                       & $\to \HOCp + \ce{Ar}$ &          & $<0.2$              \\
    \hline\hline
    $\ce{C+}+\ce{H2O}$ & $\to \HCOp + \ce{H}$  & $77(6)^c$  & 0.32$^c$               \\
                       & $\to \HOCp + \ce{H}$  &            & 0.68$^c$                \\
  \end{tabular}
  \vskip 0.2em
  \end{center}
  {\footnotesize{\emph{Note:}
  Rate coefficients in $10^{-10}\;\rateU$. 
  Errors are given in parentheses and refer to the last quoted digit. 
  Corresponding temperatures are 80~K for reactions with \ce{H2} and 125~K for reactions 
  with \ce{CO}. The values for \ce{C+ + H2O} are also given for comparison.
  \emph{a} -- \cite{Jusko2023}, \emph{b} -- \cite{Smith2002}, \emph{c} -- \cite{Yang2021}, Lang. -- Langevin rate coefficient.
  }}
\end{table}

\section{Astrochemical implications}

The use of the [\HCOp]/[\HOCp] ratio as an astronomical diagnostic tool is based on its sensitivity to the physical conditions of the environment, and, in particular, to the ionization mechanisms at play \citep{Fuente2008, Harada2021}.
This characterization relies on astrochemical models which consider two main production mechanisms for \HOCp: the reaction of \ce{C+} with \ce{H2O} and the reaction of \ce{CO+} with \ce{H2} \citep{Liszt2004}. The former reaction is known to produce \HOCp\ with a high branching ratio, while the latter reaction was long believed to produce both isomers with a 1:1 branching ratio. Our new measurements show that the reaction of \ce{CO+} with \ce{H2} does not produce significant amounts of \HOCp, effectively removing one of the main production routes for \HOCp\ in astrochemical models. 

To assess the impact of our new measured reaction rate coefficients on the predicted [\HCOp]/[\HOCp] abundance ratio, we have included our new values in the gas-grain chemical code pyRate \citep{Sipila2015} and computed the abundance and ratio of both isomers under typical conditions for a dense molecular cloud, namely a gas density of $n_{\ce{H2}} = 10^4$ cm$^{-3}$, a temperature of 10~K, and visual extinction $A_V$ of 10~mag, and a diffuse cloud, with $n_{\ce{H2}} = 5  \times 10^2$ cm$^{-3}$, a temperature of 25~K, and visual extinction $A_V$ of 1~mag. 
We focus on the effect of two particular reaction parameters: the aforementioned branching ratio of the reaction of \ce{CO+} with \ce{H2}, and the isomerization reaction of \HOCp\ to \HCOp\ with \ce{H2}, whose recommended rate coefficient in the 2024 KIDA network \citep{Wakelam2024}, $1.0 \times 10^{-11}\;\rateU$, is 40 times lower than our measured value and the previously available experimental value from \citet{Smith2002}. 

\begin{figure}[ht!]
\includegraphics[width=\linewidth]{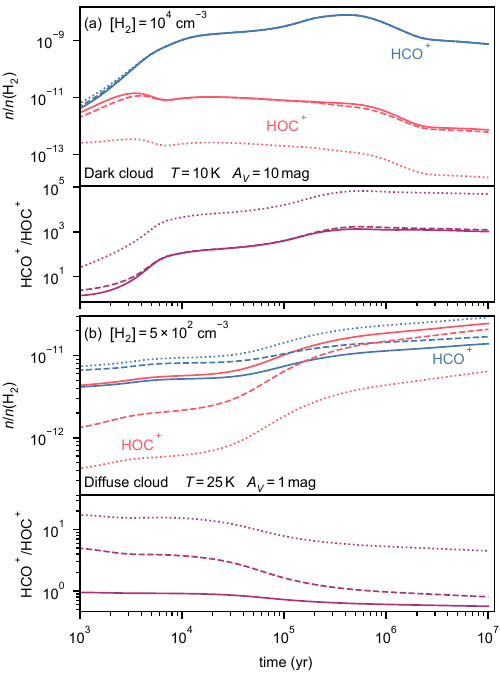}
\caption{Modeled time evolution of \HCOp, \HOCp, and their abundance ratio under typical dense cloud (upper panels) and diffuse cloud (lower panels) conditions. Solid lines: model with KIDA 2024 reaction rate coefficients and branching ratios. Dashed lines: updated branching ratio for \ce{CO+ + H2}. Dotted lines: updated branching ratio for \ce{CO+ + H2} and updated rate coefficient for the \ce{HOC+ + H2} isomerization reaction.}
\label{fig_model}
\end{figure}

The results of our modeling are shown in Fig.~\ref{fig_model}. We find that the new branching ratio for the reaction of \ce{CO+} with \ce{H2} has a significant effect on the predicted [\HCOp]/[\HOCp] ratio in diffuse clouds, increasing it by a factor of $\sim 4$ at early times (up to $\sim 10^4$ years), with the enhancement decreasing to $\sim 1.5$ at later times. In dense clouds, however, the effect is negligible. On the other hand, the new rate coefficient for the isomerization reaction of \HOCp\ to \HCOp\ with \ce{H2} has a significant effect on the predicted [\HCOp]/[\HOCp] ratio in both diffuse and dense clouds, increasing it by a factor of $\sim 100$ (for dark clouds) and $\sim 5$ (for diffuse clouds) across all timescales.

\section{Conclusions}
We have presented isomer-resolved low-temperature ion-molecule kinetics for reactions governing the chemistry of \HCOp\ and \HOCp. Our measurements show that \ce{CO+ + H2} forms almost exclusively \HCOp, with an \HOCp\ branching fraction below 0.002 under our conditions, contradicting the long-used 1:1 assumption adopted in many astrochemical networks. We also measured efficient isomerization of \HOCp\ to \HCOp\ by both \HH\ and \ce{CO}, with rate coefficients of $(4\pm1)\times10^{-10}\;\rateU$ at 80~K and $(9\pm2)\times10^{-10}\;\rateU$ at 125~K, respectively. In addition, we derived a \HOCp\ branching fraction of 0.15 $\pm$ 0.04 for \ce{H3+ + CO} at 125~K and found no significant \HOCp\ formation in the \ce{ArH+ + CO} reaction.

When incorporated into a gas-grain astrochemical model, these updated kinetics change the predicted [\HCOp]/[\HOCp] ratio substantially: strongly suppressing a commonly adopted \HOCp\ formation route via \ce{CO+ + H2} mainly affects diffuse-cloud conditions, while the larger \ce{HOC+ + H2} isomerization rate increases [\HCOp]/[\HOCp] by up to two orders of magnitude in both diffuse and dense environments. 

These new measurements also improve the interpretation of observed [\HCOp]/[\HOCp] column-density ratios across different interstellar environments. By strongly suppressing the \HOCp\ branch of the \ce{CO+ + H2} reaction and establishing an efficient \ce{HOC+ \to HCO+} isomerization rate with \HH, our results remove two major kinetic uncertainties affecting this diagnostic. Low [\HCOp]/[\HOCp] ratios observed in diffuse clouds, PDRs, and strongly irradiated environments therefore require efficient \HOCp\ formation pathways capable of competing with rapid isomerization, whereas the high ratios characteristic of dense molecular gas are naturally favored by the efficient destruction of \HOCp\ through collisions with \HH. The [\HCOp]/[\HOCp] ratio can thus provide a more quantitatively constrained probe of the balance between ionization, electron abundance, and molecular-gas density across diverse astrophysical environments.  More broadly, the revised kinetics presented here provide a substantially firmer chemical basis for exploiting \HCOp\ and \HOCp\ observations to constrain ionization-driven chemistry in the interstellar medium.

\section{Experimental methods}

The experiments were performed in a cryogenic 22-pole radio frequency ion trap, as described in detail in \cite{Jusko2023}.
The ion trapping technology allows for long storage times under controlled, variable-temperature conditions and precise sequential 
control of added neutral reactants. 
Primary ions are produced in a storage ion source \citep{Gerlich1992} using electron bombardment of a neutral precursor gas. 
The ions are trapped using collisions with a high-number-density He pulse, which guarantees translational cooling 
and relaxation of internal degrees of freedom to temperatures close to those of the trap walls.
After a variable storage time, the ions are extracted from the trap and analyzed using a quadrupole mass spectrometer with a mass resolution better than $1\;\massU$ (mass-to-charge ratio).
To complement the mass spectroscopic capability with isomer selectivity, we apply chemical probing \citep{Dohnal2023} using \ce{^{15}N2}.
For experiments involving \HOCp\ as a reactant, the ions are produced directly in the 22-pole trap from stored \ce{C+} ions in reaction with a neutral
\ce{H2O} beam (see Sec.~S1.2 in the Supporting Information, SI). This reaction is known to efficiently produce \HOCp\ \citep{Yang2021} and has also been successfully used for spectroscopic studies of \HOCp\ in the same ion trap setup
 \citep{Jimenez2024}.
Due to the highly exothermic nature of the \ce{C+ + H2O} reaction, a large fraction of the \HOCp\ ions are produced in highly excited rovibrational states, so we allow sufficient time for radiative deexcitation (tens of milliseconds) before using them in subsequent experiments.
Trap temperatures were limited by the condensation of the neutrals on the trap walls. Therefore, experiments involving \ce{CO} were performed at 125~K, while all others were performed at 80~K (limited by \iNN).
The reader is referred to the Supporting Information for a detailed description of the experimental procedures.

\subsection*{Data availability}

The data that support the findings of this study are openly available
at \url{https://doi.org/10.5281/zenodo.10663705}.

\begin{acknowledgements}
This work was supported by the Max Planck Society.
R.D. greatly acknowledges the support of the Stiftung Stipendien-Fonds of the Verband der Chemischen Industrie e.V. through its Kekulé Fellowship.
P.D. acknowledges funding from the Czech Science Foundation (GACR 23-05439S).
\end{acknowledgements}

\bibliographystyle{aa_url} 
\bibliography{lit} 

\end{document}